\documentclass[12pt]{article}
\usepackage{graphicx}
\def\bra#1{\left\langle #1\right|}
\def\ket#1{\left| #1\right\rangle}
\newcommand{\bers}{\begin{eqnarray*}}
\newcommand{\eers}{\end{eqnarray*}}
\newcommand{\bt}{\begin{itemize}}
\newcommand{\et}{\end{itemize}}
\def\beq{\begin{equation}}
\def\eeq{\end{equation}}
\def\bea{\begin{eqnarray}}
\def\eea{\end{eqnarray}}
\def\nn{\nonumber}
\def\sla#1{\raise.15ex\hbox{$/$}\kern-.57em #1}

\def\sss{\scriptscriptstyle}

\def\barp{{\raise.35ex\hbox
{${\sss (}$}}---{\raise.35ex\hbox{${\sss )}$}}}
\def\bdbarp{\hbox{$B_d$\kern-1.4em\raise1.4ex\hbox{\barp}}}
\def\bsbarp{\hbox{$B_s$\kern-1.4em\raise1.4ex\hbox{\barp}}}

\def\roughly#1{\mathrel{\raise.3ex\hbox
{$#1$\kern-.75em\lower1ex\hbox{$\sim$}}}}
\def\lsim{\roughly<}

\def\lft{{\sss L}}
\def\rht{{\sss R}}
\def\npb#1#2#3{{\it Nucl.\ Phys.} {\bf B#1} (#2) #3}
\def\plb#1#2#3{{\it Phys.\ Lett.} {\bf #1B} (#2) #3}

\def\prd#1#2#3{{\it Phys.\ Rev.} {\bf D#1} (#2) #3}
\def\newprd#1#2#3{{\it Phys.\ Rev.} {\bf D#1}: #3 (#2)}

\begin{document}

\begin{flushright}  
UdeM-GPP-TH-02-102\\
\end{flushright}
\vskip0.5truecm

\begin{center} 

{\large \bf
\centerline{New-Physics Effects on Triple-Product Correlations in
 $\Lambda_b$ Decays}}
\vspace*{1.0cm}
{\large Wafia Bensalem\footnote{email: wafia@lps.umontreal.ca},
  Alakabha Datta\footnote{email: datta@lps.umontreal.ca} 
and David
  London\footnote{email: london@lps.umontreal.ca}} \vskip0.3cm
{\it  Laboratoire Ren\'e J.-A. L\'evesque, Universit\'e de
  Montr\'eal,} \\
{\it C.P. 6128, succ.\ centre-ville, Montr\'eal, QC, Canada H3C 3J7} \\
\vskip0.5cm
\bigskip
(\today)
\vskip0.5cm
{\Large Abstract\\}
\vskip3truemm
\parbox[t]{\textwidth} {We adopt an effective-lagrangian approach to
compute the new-physics contributions to T-violating triple-product
correlations in charmless $\Lambda_b$ decays. We use factorization and
work to leading order in the heavy-quark expansion. We find that the
standard-model (SM) predictions for such correlations can be
significantly modified. For example, triple products which are
expected to vanish in the SM can be enormous ($\sim 50\%$) in the
presence of new physics. By measuring triple products in a variety of
$\Lambda_b$ decays, one can diagnose which new-physics operators are
or are not present. Our general results can be applied to any specific
model of new physics by simply calculating which operators appear in
that model.}
\end{center}
\thispagestyle{empty}
\newpage
\setcounter{page}{1}
\baselineskip=14pt

\section{Introduction}

The origin of CP violation remains one of the important open questions
in particle physics. Within the standard model (SM), CP violation is
due to the presence of phases in the Cabibbo-Kobayashi-Maskawa (CKM)
quark mixing matrix. The $B$-factories BaBar and Belle have been built
to test this: if the SM explanation is correct, we expect to observe
large CP-violating rate asymmetries in $B$ decays \cite{CPreview}. To
date, one of the CP phases of the unitarity triangle has been
measured: $\sin 2\beta = 0.78 \pm 0.08$ \cite{betameas}, which is
consistent with the SM.

Although the main focus has been on rate asymmetries, there is another
type of CP-violating signal which could potentially reveal the
presence of physics beyond the SM. Triple-product correlations of the
form $\vec v_1 \cdot (\vec v_2 \times \vec v_3)$, where each $v_i$ is
a spin or momentum, are odd under time reversal (T). Therefore, by the
CPT theorem, these are also signals of CP violation. A nonzero
triple-product correlation is signalled by a nonzero value of the
asymmetry
\beq
A_{\sss T} \equiv 
{{\Gamma (\vec v_1 \cdot (\vec v_2 \times \vec v_3)>0) - 
\Gamma (\vec v_1 \cdot (\vec v_2 \times \vec v_3)<0)} \over 
{\Gamma (\vec v_1 \cdot (\vec v_2 \times \vec v_3)>0) + 
\Gamma (\vec v_1 \cdot (\vec v_2 \times \vec v_3)<0)}} ~,
\label{Toddasym}
\eeq
where $\Gamma$ is the decay rate for the process in question. However,
there is a well-known caveat: strong phases can produce a nonzero
value of $A_{\sss T}$, even if the weak phases are zero (i.e.\ CP
violation is not really present). Thus, to be sure that one is truly
probing T and CP violation, one must compare the value of $A_{\sss T}$
with that of ${\bar A}_{\sss T}$, which is the T-odd asymmetry
measured in the CP-conjugate decay process.

Triple-product correlations can be measured in $B \to V_1 V_2$ decays,
where $V_1$ and $V_2$ are vector mesons \cite{Valencia}. In the rest
frame of the $B$, the triple product takes the form ${\vec p} \cdot
(\varepsilon_1 \times \varepsilon_2)$, where ${\vec p}$ is the
momentum of one of the final-state particles, and $\varepsilon_i$ is
the polarization of the $V_i$. One can also consider triple-product
correlations in $\Lambda_b$ decays. Since many such triple products
involve the spin of the $\Lambda_b$, this means that, in contrast to
$B$ decays, one is sensitive to the spin of the $b$-quark
\cite{BenLon}, as it is expected to provide the dominant contribution
to the spin of the $\Lambda_b$.

In a recent paper \cite{BDL}, we used factorization to study the SM
predictions for triple products in charmless two-body $\Lambda_b$
decays. We considered decays which are generated by the quark-level
transitions $b \to s \bar{q} q$ or $b \to d \bar{q} q$. These decays
take the form $\Lambda_b \to F_1 F_2$, where $F_1$ is a light
spin-$1\over 2$ baryon, such as $p$, $\Lambda$, etc., and $F_2$ is a
pseudoscalar ($P$) or vector ($V$) meson. There was only one decay in
which there was a large effect: the triple-product asymmetry for
$\Lambda_b \to p K^-$ was found to be 18\%. For all other decays, the
asymmetries are found to be at most at the percent level.

The fact that all these triple-product asymmetries are expected to be
small in the SM suggests that this is a good area to look for physics
beyond the SM. In this paper, we examine the effect of new physics on
triple products in charmless $\Lambda_b$ decays. In order to study
this, we adopt an effective-lagrangian approach: we write down all
possible dimension-six new-physics four-fermi operators at the quark
level. Then, using factorization, we compute their contributions to
the various triple-product correlations in $\Lambda_b$ decays.

There are several advantages to this approach. First, we are able to
establish which triple products can be significantly affected by the
presence of new physics. Second, we can also determine specifically
which new-physics operators contribute to these triple products. Thus,
by measuring a number of different triple-product correlations, we
may be able to diagnose which operators are or are not present.
Finally, these operators include all possible models of new physics.
Therefore one can apply our results to a specific model by simply
calculating which new-physics operators appear in that model. We will
give examples of this procedure.

The paper is organized as follows. In Section 2, we introduce the
new-physics operators used in our analysis. We also give two examples
of specific models which generate some of these operators:
supersymmetry with R-parity breaking, and $Z$- and $Z'$-mediated
flavour-changing neutral currents. We compute the contributions of the
new-physics operators to triple-product correlations in $\Lambda_b$
decays in Section 3. Here we retain only the leading term in the
heavy-quark expansion since it is very unlikely that new physics
contributes to subleading processes without affecting the
leading-order processes. In Section 4, we estimate the size of the
various triple products in the presence of new physics. By comparing
triple products in $\Lambda_b \to F_1 P$ and $\Lambda_b \to F_1 V$
decays, we examine the ``diagnostic power'' of this approach, i.e.\
the extent to which one can determine which new-physics operators are
present. We also show how our results can be applied to the specific
models of new physics discussed previously. We conclude in Section 5.

\section{New Physics}

We are interested in charmless $\Lambda_b$ decays, which are governed
by the quark-level processes $b \to s \bar{q} q$ or $b \to d \bar{q}
q$. In what follows we will concentrate on the $b\to s$ transitions;
it is straightforward to adapt our analysis to the $b\to d$ case.

Taking into account the two different colour structures, as well as
all possible Lorentz structures, there are a total of 20 dimension-six
new-physics operators which contribute to each of the $b \to s \bar{q}
q$ transitions, $q=u,d,s$. These can be written as
\bea
{\cal H}^q_{\sss NP} = \sum_{\sss A,B = L,R} {4 G_{\sss F} \over \sqrt{2}} &&
\hskip-5truemm
\left\{ 
f_{q,1}^{\sss AB} \, {\bar s}_\alpha \gamma_{\sss A} b_\beta \, {\bar
q}_\beta \gamma_{\sss B} q_\alpha + f_{q,2}^{\sss AB} \, {\bar s}
\gamma_{\sss A} b \, {\bar q} \gamma_{\sss B} q \right. \nn\\
&& \hskip-7truemm
+ ~ g_{q,1}^{\sss AB} \, {\bar s}_\alpha \gamma^\mu \gamma_{\sss A}
b_\beta \, {\bar q}_\beta \gamma_\mu \gamma_{\sss B} q_\alpha +
g_{q,2}^{\sss AB} \, {\bar s} \gamma^\mu \gamma_{\sss A} b \, {\bar q}
\gamma_\mu \gamma_{\sss B} q \nn\\
&& \hskip-7truemm \left.
+ ~ h_{q,1}^{\sss AB} \, {\bar s}_\alpha \sigma^{\mu\nu} \gamma_{\sss
A} b_\beta \, {\bar q}_\beta \sigma_{\mu\nu} \gamma_{\sss B} q_\alpha
+ h_{q,2}^{\sss AB} \, {\bar s} \sigma^{\mu\nu} \gamma_{\sss A} b \,
{\bar q} \sigma_{\mu\nu} \gamma_{\sss B} q \right\},
\label{HeffNP}
\eea
where we have defined $\gamma_{\sss R(L)}= {1\over 2} (1 \pm
\gamma_5)$. Note: although we have written the tensor operators in the
same compact form as the other operators, it should be noted that
those with $\gamma_{\sss A} \ne \gamma_{\sss B}$ are identically
zero. Thus, one can effectively set $h_{q,i}^{\sss LR} = h_{q,i}^{\sss
RL} = 0$.

All models of new physics which contribute to $b \to s \bar{q} q$ will
generate operators found in the above effective hamiltonian. These can
arise at tree level (e.g.\ supersymmetry with R-parity breaking, $Z$-
and $Z'$-mediated flavour-changing neutral currents, models with
flavour-changing neutral scalars, etc.) or at loop level (e.g.\
minimal supersymmetry, left-right symmetric models, four generations,
etc.) \cite{Bnewphysics}. In some cases one will obtain operators of
the form ${\bar q} {\cal O} b \, {\bar s} {\cal O}' q$, but one can
perform a Fierz transformation to put them into the form of
Eq.~(\ref{HeffNP}). Note that, in general, models of new physics do
not lead directly to tensor operators ($h_{q,i}^{\sss AB}$), since
typically only vector or scalar particles are involved. However, such
tensor operators can arise when other operators are Fierz-transformed
into the above form, so they must be included in our analysis (the
scalar operator ${\bar q} b {\bar s} q$ is such an example).

Because the new-physics operators are of dimension six, by dimensional
analysis we expect them to be suppressed by a factor $\Lambda^2$,
where $\Lambda$ is the scale of new physics. However, with the
normalization in Eq.~(\ref{HeffNP}), the suppression factor is only
$M_{\sss W}^2$. We therefore expect the size of the coefficients
$f_{q,i}^{\sss AB}$, $g_{q,i}^{\sss AB}$ and $h_{q,i}^{\sss AB}$ to be
naturally of $O(M_{\sss W}^2/\Lambda^2) \sim 10^{-2}$ for a
new-physics scale of about 1 TeV.

Even so, these new-physics effects can be quite significant. In the
SM, one finds only operators of the form ${\bar s} \gamma^\mu
\gamma_\lft b \, {\bar q} \gamma_\mu \gamma_{\sss L,R} q$, with both
colour assignments. These operators are typically multiplied by one of
two factors: either (i) the CKM matrix elements $V_{tb} V_{ts}^*$
times a Wilson coefficient of $O(10^{-2})$, or (ii) $V_{ub} V_{us}^*$
times a Wilson coefficient of $O(1)$. In either case, new-physics
operators with coefficients of $O(10^{-2})$ would actually {\it
dominate} over the SM contributions. (This is, in part, what allows us
to put constraints on specific models of new physics.) The bottom line
is that the new operators of Eq.~(\ref{HeffNP}) can contribute
substantially to charmless $\Lambda_b$ decays.

As noted above, by construction the effective hamiltonian of
Eq.~(\ref{HeffNP}) includes all possible models of new physics. Of
course, in a particular new-physics model, only a subset of the new
operators will appear. Our general analysis can then be applied to
that specific model by retaining only the coefficients of the nonzero
operators. In order to show explicitly how this works, below we give
two examples of such specific models.

\subsection{Supersymmetry with R-parity breaking}

In supersymmetric models, the $R$-parity of a field with spin $S$,
baryon number $B$ and lepton number $L$ is defined to be
\beq
R=(-1)^{2S+3B+L} ~.
\eeq
$R$ is $+1$ for all the SM particles and $-1$ for all the
supersymmetric particles. $R$-parity invariance is often imposed on
the Lagrangian in order to maintain the separate conservation of
baryon number and lepton number. Imposition of $R$-parity conservation
has some important consequences: super particles must be produced in
pairs in collider experiments and the lightest super particle (LSP)
must be absolutely stable. The LSP therefore provides a good candidate
for cold dark matter.

Despite the above-mentioned attractive features of R-parity
conservation, this conservation is not dictated by any fundamental
principle such as gauge invariance, so that there is no compelling
theoretical motivation for it. The most general superpotential of the
MSSM, consistent with $SU(3)\times SU(2)\times U(1)$ gauge symmetry
and supersymmetry, can be written as
\beq
{\cal W}={\cal W}_R+{\cal W}_{\not \! R}~,
\eeq
where ${\cal W}_R$ is the $R$-parity conserving piece, and ${\cal
W}_{\not \! R}$ breaks $R$-parity. They are given by
\bea
{\cal W}_R&=&h_{ij}L_iH_2E_j^c+h_{ij}^{\prime}Q_iH_2D_j^c
             +h_{ij}^{\prime\prime}Q_iH_1U_j^c ~,\\ \label{RV}
{\cal W}_{\not \! R}&=&\frac{1}{2}\lambda_{[ij]k}L_iL_jE_k^c
+\lambda_{ijk}^{\prime}L_iQ_jD_k^c
             +\frac{1}{2}\lambda_{i[jk]}^{\prime\prime}
U_i^cD_j^cD_k^c+\mu_iL_iH_2 ~.
\label{lag}
\eea
Here $L_i(Q_i)$ and $E_i(U_i,D_i)$ are the left-handed lepton (quark)
doublet and lepton (quark) singlet chiral superfields, where $i,j,k$
are generation indices and $c$ denotes a charge conjugate field.
$H_{1,2}$ are the chiral superfields representing the two Higgs
doublets.

In the $R$-parity-violating superpotential [Eq.~(\ref{RV})], the
$\lambda$ and $\lambda^{\prime}$ couplings violate lepton number
conservation, while the $\lambda^{\prime\prime}$ couplings violate
baryon number conservation.  $\lambda_{[ij]k}$ is antisymmetric in the
first two indices and $\lambda^{\prime\prime}_{i[jk]}$ is
antisymmetric in the last two indices.  There are therefore 27
$\lambda^{\prime}$-type couplings and 9 each of the $\lambda$ and
$\lambda^{\prime \prime}$ couplings.  While it is theoretically
possible to have both baryon-number and lepton-number violating terms
in the Lagrangian, the non-observation of proton decay imposes very
stringent conditions on their simultaneous presence \cite{Proton}.
One therefore assumes the existence of either $L$-violating couplings
or $B$-violating couplings, but not both. The terms proportional to
$\lambda$ are not relevant to our present discussion and will not be
considered further.

We begin with the $B$-violating couplings. The transition $b \to s
\bar{u} u$ can be generated at tree level through the t-channel
exchange of the $d$-squark, $\tilde{d_R}$, with strength proportional
to $|\lambda^{\prime\prime}_{112} \lambda^{*\prime\prime}_{113}|$.
However, this product of couplings is already constrained to be $\sim
10^{-8}$ from $n-{\bar{n}}$ oscillations and double nucleon decay
\cite{Rpreview}. There are therefore no significant contributions to
the new-physics operators of Eq.~(\ref{HeffNP}) corresponding to
$q=u$.

Similarly, the antisymmetry of the $B$-violating couplings,
$\lambda^{\prime\prime}_{i[jk]}$ in the last two indices implies that
there are no operators that can generate the $b \to s \bar{s} s$
transition, so that all the operators in Eq.~(\ref{HeffNP}) vanish for
$q=s$.

Finally, the operators that generate the $b \to s \bar{d} d$
transition are given by \cite{He}
\beq
L_{eff} = \frac{\lambda^{\prime \prime}_{i12} \lambda^{\prime
\prime*}_{i13}} {4 m_{ \widetilde{u}_i}^2} ( \bar d_\alpha
\gamma_\mu\gamma_ R d_\alpha \, \bar s_\beta \gamma_\mu \gamma_R
b_\beta -\bar d_\alpha \gamma_\mu \gamma_R d_\beta \, \bar s_\beta
\gamma_\mu \gamma_R b_\alpha) ~.
\eeq
Hence the only nonvanishing operators in Eq.~(\ref{HeffNP}) are
\beq
g_{d,1}^{\sss RR}=-g_{d,2}^{\sss RR}=
-\frac{\sqrt{2}}{G_F}\frac{\lambda^{\prime \prime}_{i12} 
\lambda^{\prime \prime*}_{i13}}
{16  m_{ \widetilde{u}_i}^2} ~.
\label{Bsdd}
\eeq
As mentioned above, the constraint on $|{\lambda^{\prime \prime}_{112}
\lambda^{\prime \prime*}_{113}}|$ is at the $10^{-8}$ level. However,
the constraint on $|{\lambda^{\prime \prime}_{i12} \lambda^{\prime
\prime*}_{i13}}|$, $i \ne 1$, comes only from the nonleptonic decay
$B^- \to \bar{K^0} \pi^{-}$ \cite{He}, and is much weaker:
\beq
|{\lambda^{\prime \prime}_{i12} \lambda^{\prime \prime*}_{i13}}|( i
\ne 1) \le  1.03 \times 10^{-2} ~,
\eeq
where a squark mass $m_{\widetilde{f}}=100$ GeV has been assumed. We
therefore find
\beq
|g_{d,1}^{\sss RR}| = |g_{d,2}^{\sss RR}| \le 7.6 \times 10^{-3} ~.
\label{Bsddn}
\eeq

We now turn to the $L$-violating couplings. In terms of four-component 
Dirac spinors, these are given by \cite{DatXin}
\bea
{\cal L}_{\lambda^{\prime}}&=&-\lambda^{\prime}_{ijk}
\left [\tilde \nu^i_L\bar d^k_R d^j_L+\tilde d^j_L\bar d^k_R\nu^i_L
       +(\tilde d^k_R)^*(\bar \nu^i_L)^c d^j_L\right.\nonumber\\
& &\hspace{1.5cm} \left. -\tilde e^i_L\bar d^k_R u^j_L
       -\tilde u^j_L\bar d^k_R e^i_L
       -(\tilde d^k_R)^*(\bar e^i_L)^c u^j_L\right ]+h.c.\
\label{Lviolating}
\eea
There are a variety of sources which bound the above couplings
\cite{Rpreview, He}. For the sake of brevity we will only quote the
bounds and not their sources. Assuming a commom sfermion mass of 100
GeV we find the most stringent bounds are
\bea
& |{\lambda^{\prime}_{i12} \lambda^{ \prime*}_{i13}}|( i \ne 1) \le
1.7 \times 10^{-3} ~~,~~ 
|{\lambda^{\prime}_{112} \lambda^{ \prime*}_{113}}| \le 4.4 \times
10^{-4} & \\
& |{\lambda^{\prime}_{111*} \lambda^{ \prime*}_{132}}| \le 1.4 \times
10^{-4} ~~,~~
|{\lambda^{\prime*}_{211} \lambda^{ \prime}_{232}}| \le 4.7 \times
10^{-4} ~~,~~
|{\lambda^{\prime}_{311*} \lambda^{ \prime}_{332}}| \le 4.7 \times
10^{-4} & \nn\\
& |{\lambda^{\prime}_{111} \lambda^{ \prime*}_{123}}| \le 2.2 \times
10^{-5} ~~,~~
|{\lambda^{\prime}_{211} \lambda^{ \prime*}_{223}}| \le 2.2\times
10^{-3} ~~,~~
|{\lambda^{\prime}_{311} \lambda^{ \prime*}_{323}}| \le 2.2\times
10^{-3} & \nn\\
& |{\lambda^{\prime}_{131} \lambda^{ \prime*}_{121}}| \le 8.2 \times
10^{-4} ~~,~~
|{\lambda^{\prime}_{231} \lambda^{ \prime*}_{221}}| \le 1.3\times
10^{-3} ~~,~~
|{\lambda^{\prime}_{331} \lambda^{ \prime*}_{321}}| \le 1.3\times
10^{-3} & \nn\\
& |{\lambda^{\prime}_{132} \lambda^{ \prime*}_{122}}| \le 1.2 \times
10^{-2} ~~,~~
|{\lambda^{\prime}_{232} \lambda^{ \prime*}_{222}}| \le 1.2 \times
10^{-1} ~~,~~
|{\lambda^{\prime}_{332} \lambda^{ \prime*}_{322}}| \le 2.3 \times
10^{-1} & \nn\\
& |{\lambda^{\prime}_{122} \lambda^{ \prime*}_{123}}| \le 1.8 \times
10^{-3} ~~,~~
|{\lambda^{\prime}_{223} \lambda^{ \prime*}_{222}}| \le 4.4 \times
10^{-2} ~~,~~
|{\lambda^{\prime}_{322} \lambda^{ \prime*}_{323}}| \le 2.7 \times
10^{-1} ~. & \nn
\label{Lviolbounds}
\eea

There is a single contribution to the $b\to s {\bar{u}} u$ transition:
\beq
L_{eff}=
-\frac{\lambda^{\prime}_{i12} \lambda^{\prime*}_{i13}}
{2  m_{ \widetilde{e}_i}^2}
\bar u_\alpha \gamma_\mu \gamma_L u_\beta \,
\bar s_\beta \gamma_\mu \gamma_R b_\alpha ~.
\eeq
Hence the only nonvanishing operator for $q=u$ in Eq.~(\ref{HeffNP}) is
\beq
g_{u,1}^{\sss RL}= -\frac{\sqrt{2}}{G_F}\frac{\lambda^{\prime }_{i12}
\lambda^{\prime*}_{i13}} {8 m_{ \widetilde{e}_i}^2} ~.
\label{BsuuL}
\eeq
Using the bounds of Eq.~(\ref{Lviolbounds}), we find
\beq
|g_{u,1}^{\sss RL}| \le 2.6 \times 10^{-3} ~.
\eeq

Turning now to the $b \to s {\bar{d}}d $ transition the
relevant Lagrangian is
\bea
L_{eff}&=&
\frac{\lambda^{\prime}_{i11} \lambda^{\prime*}_{i23}}{  m_{ \widetilde{\nu}_i}^2}
\bar d \gamma_L d_\beta \,
\bar {s}\gamma_R b
+\frac{\lambda^{\prime}_{i32} 
\lambda^{\prime*}_{i11}}{  m_{ \widetilde{\nu}_i}^2}
\bar d \gamma_R d \, \bar s \gamma_L b \nonumber \\
&-&\frac{\lambda^{\prime}_{i12} \lambda^{\prime*}_{i13}}
{2  m_{ \widetilde{\nu}_i}^2}
\bar d_\alpha \gamma_\mu \gamma_L d_\beta \, \bar s_\beta \gamma_\mu 
\gamma_R b_\alpha
-\frac{\lambda^{\prime}_{i31} \lambda^{\prime *}_{i21}}
{2  m_{ \widetilde{\nu}_i}^2}
\bar d_\alpha \gamma_\mu \gamma_R d_\beta \, \bar s_\beta \gamma_\mu \gamma_L 
b_\alpha ~.
\label{Lbsdd}
\eea
The nonvanishing operators in Eq. \ref{HeffNP} are then
\[
f_{d,2}^{\sss LR} = \frac{\sqrt{2}}{G_F} \frac{\lambda^{\prime}_{i32}
\lambda^{\prime*}_{i11}} {4 m_{ \widetilde{\nu}_i}^2} ~~,~~
f_{d,2}^{\sss RL} = \frac{\sqrt{2}}{G_F} \frac{\lambda^{\prime}_{i11}
\lambda^{\prime*}_{i23}} {4 m_{ \widetilde{\nu}_i}^2} ~, 
\]
\vskip-2truemm
\beq
g_{d,1}^{\sss RL} = g_{u,1}^{\sss RL*} ~~,~~
g_{d,1}^{\sss LR} = -\frac{\sqrt{2}}{G_F} \frac{\lambda^{\prime}_{i31}
\lambda^{\prime*}_{i21}} {8 m_{ \widetilde{\nu}_i}^2} ~,
\label{BsddL}
\eeq
%
with
\bea
& |f_{d,2}^{\sss LR}| \le 1.4 \times 10^{-3} ~~,~~
 |f_{d,2}^{\sss RL}| \le 6.6 \times 10^{-3} ~, & \nn\\
& |g_{d,1}^{\sss RL}| \le 2.6 \times 10^{-3} ~~,~~
 |g_{d,1}^{\sss LR}| \le 2.0 \times 10^{-3} ~. & 
\eea

Finally, turning to the $b \to s {\bar{s}}s $ transition, the relevant
Lagrangian is
\beq
L_{eff} = \frac{\lambda^{\prime}_{i32} \lambda^{\prime*}_{i22}} { m_{
\widetilde{\nu}_i}^2} \bar s \gamma_R s \, \bar {s}\gamma_L b+
\frac{\lambda^{\prime}_{i22} \lambda^{\prime*}_{i23}} { m_{
\widetilde{\nu}_i}^2} \bar s \gamma_L s \, \bar {s}\gamma_R b ~,
\eeq
allowing the identification
\beq
f_{s,2}^{\sss LR} = \frac{\sqrt{2}}{G_F} \frac{\lambda^{\prime}_{i32}
\lambda^{\prime*}_{i22}} {4 m_{ \widetilde{\nu}_i}^2} ~~,~~
f_{s,2}^{\sss RL} = \frac{\sqrt{2}}{G_F} \frac{\lambda^{\prime}_{i22}
\lambda^{\prime*}_{i23}} {4 m_{ \widetilde{\nu}_i}^2} ~,
\label{BsssL}
\eeq
with
\beq
|f_{s,2}^{\sss LR}| \le 0.7 ~~,~~ |f_{s,2}^{\sss RL}| \le 0.8 ~.
\eeq

\subsection{$Z$- and $Z'$-mediated FCNC's}

In these models, one introduces an additional vector-singlet charge
$-1/3$ quark $h$, as is found in $E_6$ grand unified theories, and
allows it to mix with the ordinary down-type quarks $d$, $s$ and
$b$. Since the weak isospin of the exotic quark is different from that
of the ordinary quarks, flavor-changing neutral currents (FCNC's)
involving the $Z$ are induced \cite{NirSilv}. The $Zb{\bar s}$ FCNC
coupling, which is of interest to us here, is parametrized by the
independent parameter $U_{sb}^{\sss Z}$:
\beq
{\cal L}^{\sss Z}_{\sss FCNC} = - {g \over 2 \cos\theta_{\sss W}} \,
U_{sb}^{\sss Z} \, {\bar s}_\lft \gamma^\mu b_\lft Z_\mu ~.
\eeq
Note that it is only the mixing between the left-handed components of
the ordinary and exotic quarks which is responsible for the FCNC:
since $s_\rht$, $b_\rht$ and $h_\rht$ all have the same $SU(2)_\lft
\times U(1)_{\sss Y}$ quantum numbers, their mixing cannot generate
flavour-changing couplings of the $Z$. Models with $Z$-mediated FCNC's
will therefore generate the $g_{q,2}^{\sss LL}$ and $g_{q,2}^{\sss
LR}$ new-physics operators Eq.~(\ref{HeffNP}). These are the same
operators that appear in the SM, so that this model does not generate
new operators. (That is, these are effectively new contributions to
the electroweak penguin operators of the SM.)

The strongest constraint on $U_{sb}^{\sss Z}$ comes from the
measurement of $B(B\to \ell^+ \ell^- X)$. The most recent result from
BELLE gives \cite{BELLE}
\beq
B(B \to X_s e^+ e^-) \le 1.01 \times 10^{-5} ~,
\label{BELLEbound}
\eeq
leading to the constraint
\beq
\left\vert U_{sb}^{\sss Z} \right\vert \le 7.6 \times 10^{-4} ~.
\label{ZFCNCconstraint}
\eeq
With this constraint, it is straightforward to compute the maximal
size of the couplings $g_{q,2}^{\sss LL}$ and $g_{q,2}^{\sss LR}$. We
find
\bea
|g_{u,2}^{\sss LL}| \le 2.7 \times 10^{-4} &,& |g_{u,2}^{\sss LR}| \le
1.1 \times 10^{-4} ~, \nn\\
|g_{d,2}^{\sss LL}| = |g_{s,2}^{\sss LL}| \le 3.2 \times 10^{-4} &,& 
|g_{d,2}^{\sss LR}| = |g_{s,2}^{\sss LR}| \le 6.1 \times 10^{-5} ~.
\label{ZFCNCcoeffs}
\eea
These couplings are therefore comparable in size to those of the SM.

Of course, since no new operators are generated in this scenario, and
since the new-physics effects are about the same size as in the SM,
one does not expect large deviations from the SM predictions due to
$Z$-mediated FCNC's. However, models of new physics which contain
exotic fermions also predict, in general, the existence of additional
neutral $Z'$ gauge bosons. If the $s$-, $b$- and $h$-quarks have
different quantum numbers under the new $U(1)$ symmetry, their mixing
will induce FCNC's due to $Z'$ exchange \cite{Z'FCNC}. 

In general, as was the case for $Z$-mediated FCNC's, such
flavour-changing couplings will be constrained by the measurement of
$B(B\to \ell^+ \ell^- X)$. However, if the $Z'$ is leptophobic, i.e.\
it does not couple to charged leptons, one can evade the constraints
due to Eq.~(\ref{BELLEbound}). Such models were considered in
Ref.~\cite{LerLon}. In this case, it is the mixing of the right-handed
components of the ordinary and exotic quarks which is most important,
and we parametrize the flavour-changing $Z' b{\bar s}$ coupling as
\beq 
{\cal L}^{\sss Z'}_{\sss FCNC} = - {g \over 2 \cos\theta_{\sss W}} \,
U_{sb}^{\sss Z'} \, {\bar s}_\rht \gamma^\mu b_\rht Z'_\mu ~.
\eeq
Thus, these models will generate new operators. In particular, the
coefficients $g_{q,2}^{\sss RL}$ and $g_{q,2}^{\sss RR}$ will be
nonzero.

Even though the $Z'$ is leptophobic, there are constraints on $U^{\sss
Z'}_{sb}$ coming from the ALEPH limit $B(b \to s \nu {\bar \nu}) \le
6.4 \times 10^{-4}$ \cite{bsnunu}. In addition, in realistic models,
leptophobia is realized only approximately -- there will always be
threshold effects which produce a small coupling of the $Z'$ to
charged leptons, in which case there are constraints from
Eq.~(\ref{BELLEbound}). The constraints from both of these sources
turn out to be similar in size, and lead to \cite{LerLon}
\beq
\left| U^{\sss Z'}_{sb} \right| {M_{\sss Z}^2 \over M_{\sss Z'}^2}
\lsim 6 \times 10^{-3} ~.
\label{Z'constraint2}
\eeq
With this constraint, one can estimate how large the new-physics
coefficients can be. One finds
\beq
|g_{q,2}^{\sss AB}| \le (1-2) \times 10^{-3} ~, AB = RR, RL ~, q =
 u,d,s ~,
\label{Z'FCNCcoeffs}
\eeq
which is about an order of magnitude larger than the coefficients of
Eq.~(\ref{ZFCNCcoeffs}). Thus, $Z'$-mediated FCNC's can contribute
significantly to charmless hadronic $\Lambda_b$ decays, and can lead
to significant deviations from the SM predictions for triple products
in such processes.

\section{Triple Products}

In this section, we compute the contributions from the new-physics
operators to triple-product correlations in $\Lambda_b$ decays. In all
cases, we retain only the leading term in the heavy-quark expansion,
and neglect terms of order $m/m_{\Lambda_b}$, where $m$ is the mass of
the light meson. The main reason is that it is very unlikely that new
physics will contribute at subleading order, but not at leading
order. Indeed, as we will see, this situation can arise only in
fine-tuned scenarios. A secondary reason is that the subleading terms
are quite a bit smaller, e.g.\ $m_{K^*}/m_{\Lambda_b} \sim 15\%$.

We begin with a review of the results of the SM.

\subsection{SM Results}

In this subsection, we summarize the predictions of the SM for triple
products in $\Lambda_b$ decays. The discussion is somewhat cursory,
and we refer to the reader to Ref.~\cite{BDL} for more details.

We first consider the decay $\Lambda_b\to F_1 P$, whose amplitude can
be written generally as
\beq
{\cal M}_{\sss P}= A(\Lambda_b\to F_1 P) = i {\bar u}_{\sss F_1} (a +
b\gamma_{5}) u_{\Lambda_b} ~.
\label{pscalar}
\eeq
The calculation of $|{\cal M}_{\sss P}|^2$ yields a single triple-product
term:
\beq
{\rm Im}(a b^*) \epsilon_{\mu\nu\rho\sigma} p^\mu_{\sss F_1} s^\nu_{\sss F_1}
p^\rho_{\Lambda_b} s^\sigma_{\Lambda_b} ~,
\label{scalarTP}
\eeq
where $p^\mu_i$ and $s^\mu_i$ are the 4-momentum and polarization of
particle $i$. In the rest frame of the $\Lambda_b$, this takes the
form ${\vec p}_{\sss F_1} \cdot ({\vec s}_{\sss F_1} \times {\vec
s}_{\Lambda_b})$.

Within factorization, one can write
\bea
A(\Lambda_b\to F_1 P) & = & \sum_{O,O'} \bra{P} O \ket{0} \, \bra{F_1}
O' \ket{\Lambda_b} \nn\\
& = & i f_{\sss P} q^{\mu} \bra{F_1} \bar{q}_1 \gamma_\mu(1-\gamma_5)
b\ket{\Lambda_b} X_{\sss P} \nn\\
& & \hskip10truemm
+ i f_{\sss P} q^{\mu} \bra{F_1} \bar{q}_1
\gamma_\mu(1+\gamma_5) b\ket{\Lambda_b} Y_{\sss P} ~,
\label{Pfacformulas}
\eea
where we have defined the pseudoscalar decay constant $f_{\sss P}$ as
\beq
i f_{\sss P} q^{\mu}= \bra{P} \bar{q}_2 \gamma^{\mu}(1 - \gamma_5) q_3 \ket{0} ~,
\eeq
where $q^\mu \equiv p^\mu_{\Lambda_b} - p^\mu_{\sss F_1}$ is the
four-momentum transfer. The key point here is that, in order to obtain
a nonzero triple-product correlation, one must have two interfering
amplitudes, i.e.\ $X_{\sss P}$ and $Y_{\sss P}$ must both be nonzero, and must have
a relative weak phase. Furthermore, the triple product will be large
only if $X_{\sss P}$ and $Y_{\sss P}$ are of similar size.

One can also show that the parameters $a$ and $b$ of
Eq.~(\ref{pscalar}) can be written as
\bea
a & = & f_{\sss P} (X_{\sss P}+Y_{\sss P}) (m_{\Lambda_b}-m_{F_1}) f_1 ~,
\nn\\
b & = & f_{\sss P} (X_{\sss P}-Y_{\sss P}) (m_{\Lambda_b}+m_{F_1}) g_1 ~,
\label{abTP}
\eea
where we have dropped terms of $O(m_{\sss P}/m_{\Lambda_b})$, and
$f_1$ and $g_1$ are Lorentz-invariant form factors:
\bea
\bra{F_1} \bar{q}_1 \gamma^\mu b \ket{\Lambda_b} & =& \bar{u}_{\sss F_1}
\left[ f_1 \gamma^\mu + i \frac{f_2}{m_{\Lambda_b}}\sigma^{\mu\nu}
q_\nu + \frac{f_3}{m_{\Lambda_b}} q^\mu \right] u_{\Lambda_b} \nn \\
\bra{F_1} \bar{q}_1 \gamma^\mu \gamma_5 b \ket{\Lambda_b} & =&
\bar{u}_{\sss F_1} \left[ g_1 \gamma^\mu + i
\frac{g_2}{m_{\Lambda_b}}\sigma^{\mu\nu} q_\nu +
\frac{g_3}{m_{\Lambda_b}} q^\mu \right] \gamma_5 u_{\Lambda_b} ~.
\label{ffactor}
\eea
{}From Eqs.~(\ref{scalarTP}) and (\ref{abTP}), we therefore see
explicitly that the triple product in $\Lambda_b\to F_1 P$ is
proportional to ${\rm Im}(a b^*) \sim {\rm Im}(X_{\sss P} Y_{\sss
P}^*)$.

In the SM, there is only one class of decays which is expected to show
a significant effect \cite{BDL}: the triple-product correlation for
$\Lambda_b \to p K^-$ is found to be $\sim 18\%$. For decays such as
$\Lambda_b \to \Lambda \eta$, $\Lambda_b \to \Lambda \eta'$ and
$\Lambda_b \to n {\bar K}^0$, the triple product is less than 1\%. The
fundamental reason for this is that $\Lambda_b \to p K^-$ is governed
by the quark-level transition $b\to s {\bar u} u$, which has both a
tree and a penguin contribution, whereas the other decays are
dominated by the $b\to s$ penguin amplitude. Thus, for $\Lambda_b \to
\Lambda \eta, \Lambda \eta', n {\bar K}^0$, there is essentially only
a single decay amplitude, which precludes any CP- and T-violating
effects.

In the above discussion, the size of the triple products has been
estimated within factorization. However, it is well-known that
nonfactorizable effects can be important in $\Lambda_c$ decays. For
example, the decay $\Lambda_c \to \Sigma^+ \phi$ has been observed
\cite{nonfacdecay}, and this can only proceed via a (nonfactorizable)
$W$-exchange diagram. This then begs the question of whether
nonfactorizable effects might be important in $\Lambda_b$ decays. In
fact, the answer is that $\Lambda_b$ decays are {\it not} expected to
be significantly affected by such effects. In Ref.~\cite{Cheng}, it
was found that the $W$-exchange contributions to inclusive $\Lambda_b$
decays are suppressed relative to those in $\Lambda_c$ decays by
$O(m_c/m_b)^3$. This implies that even for exclusive decays such
nonfactorizable $W$-exchange terms are expected to be small. This was
confirmed in Ref.~\cite{BDL}: the $W$-exchange contributions to
$\Lambda_b \to p K^-$ were estimated using a pole model, and the ratio
of nonfactorizable to factorizable contributions was found to be
tiny. For these reasons, here and below we ignore all nonfactorizable
effects in $\Lambda_b$ decays.

Turning to $\Lambda_b\to F_1 V$, the general decay amplitude can be
written as \cite{SPT}
\beq
{\cal M}_{\sss V}= Amp(\Lambda_{\sss F_1}\to B V) = {\bar u}_{\sss F_1}
\varepsilon^*_\mu \left[ (p_{\Lambda_b}^\mu + p_{\sss F_1}^\mu)
(a+b\gamma_{5}) + \gamma^{\mu} (x+y\gamma_{5}) \right] u_{\Lambda_b}
~,
\label{vector}
\eeq
where $\varepsilon^*_\mu$ is the polarization of the vector meson. In
calculating $|{\cal M}_{\sss V}|^2$, one finds many triple-product terms:
\bea
|{\cal M}_{\sss V}|^2_{t.p.} & = & 2 \, {\rm Im}(a b^*) \left\vert
\varepsilon_{\sss V} \cdot (p_{\Lambda_b} + p_{\sss F_1})
\right\vert^2 \, \epsilon_{\mu\nu\rho\sigma} p^\mu_{\sss F_1}
s^\nu_{\sss F_1} p^\rho_{\Lambda_b} s^\sigma_{\Lambda_b} \nn\\
&& + 2 \, {\rm Im} \left(x y^*\right) \epsilon_{\alpha\beta\mu\nu}
\left[ \varepsilon_{\sss V} \cdot s_{\sss F_1} p_{\sss F_1}^\alpha
p_{\Lambda_b}^\beta s_{\Lambda_b}^\mu \varepsilon_{\sss V}^\nu -
\varepsilon_{\sss V} \cdot p_{\sss F_1} s_{\sss F_1}^\alpha
p_{\Lambda_b}^\beta s_{\Lambda_b}^\mu \varepsilon_{\sss V}^\nu
\right. \nn\\
&& \hskip1.2truein \left. + \varepsilon_{\sss V} \cdot s_{\Lambda_b}
p_{\sss F_1}^\alpha s_{\sss F_1}^\beta \varepsilon_{\sss V}^\mu
p_{\Lambda_b}^\nu - \varepsilon_{\sss V} \cdot p_{\Lambda_b} p_{\sss
F_1}^\alpha s_{\sss F_1}^\beta \varepsilon_{\sss V}^\mu
s_{\Lambda_b}^\nu \right] \nn\\
&& + 2 \, \varepsilon_{\sss V} \cdot \left( p_{\Lambda_b} + p_{\sss
F_1} \right) \epsilon_{\alpha\beta\mu\nu} \left[ {\rm Im} \left( a x^*
+ b y^* \right) p_{\sss F_1}^\alpha s_{\sss F_1}^\beta
p_{\Lambda_b}^\mu \varepsilon_{\sss V}^\nu \right. \nn\\
&& \hskip1.8truein + m_{\Lambda_b} {\rm Im} \left( b x^* + a y^*
\right) p_{\sss F_1}^\alpha s_{\sss F_1}^\beta s_{\Lambda_b}^\mu
\varepsilon_{\sss V}^\nu \nn\\
&& \hskip1.8truein - {\rm Im} \left( a x^* - b y^* \right) p_{\sss
F_1}^\alpha p_{\Lambda_b}^\beta s_{\Lambda_b}^\mu \varepsilon_{\sss
V}^\nu \nn\\ && \hskip1.8truein \left. - m_{\sss F_1} {\rm Im} \left(
a y^* - b x^* \right) s_{\sss F_1}^\alpha p_{\Lambda_b}^\beta
s_{\Lambda_b}^\mu \varepsilon_{\sss V}^\nu \right] ~.
\label{vecTPs}
\eea

Similar to $\Lambda_b\to F_1 P$ decays, using factorization, one can
write
\bea
\label{Vamps}
A(\Lambda_b\to  F_1 V) &=& m_{\sss V} g_{\sss V} \left\{
\varepsilon_{\mu}^{*} \bra{F_1} \bar{q}_1 \gamma^\mu(1-\gamma_5)
b\ket{\Lambda_b} X_{\sss V} \right. \nn\\
&& \qquad\qquad + \, \varepsilon_{\mu}^{*} \bra{F_1}\bar{q}_1
\gamma^\mu(1+\gamma_5) b\ket{\Lambda_b} Y_{\sss V} \\
&& \qquad\qquad + \, \varepsilon \cdot (p_{\Lambda_b} + p_{\sss F_1})
\, q_\mu \, \bra{F_1}\bar{q}_1 \gamma^\mu(1-\gamma_5) b\ket{\Lambda_b}
\frac{A_{\sss V}}{m_{\Lambda_b}^2}\nn\\
&& \qquad\qquad \left. + \, \varepsilon \cdot (p_{\Lambda_b} + p_{\sss
F_1}) \, q_\mu \, \bra{F_1}\bar{q}_1 \gamma^\mu(1+\gamma_5)
b\ket{\Lambda_b} \frac{B_{\sss V}}{m_{\Lambda_b}^2} \right\}~, \nn
\eea
where the decay constant $g_{\sss V}$ has been defined as
\beq
m_{\sss V} g_{\sss V} \varepsilon_{\mu}^{*} = \bra{V} \bar{q}_2
\gamma_{\mu} q_3 \ket{0} ~.
\eeq
(Above we have included explicit factors of $m_{\Lambda_b}^2$ so that
$A_{\sss V}$ and $B_{\sss V}$ have the same dimensions as $X_{\sss V}$
and $Y_{\sss V}$. This differs from Ref.~\cite{BDL}. Also, note that,
since the magnitudes of $p_{\Lambda_b}$, $p_{\sss F_1}$ and $q$ are
all of order $m_{\Lambda_b}$, the $A_{\sss V}$ and $B_{\sss V}$
operators are not apriori smaller than the $X_{\sss V}$ and $Y_{\sss
V}$ operators.)  Hence, using factorization, the quantities $a$, $b$,
$x$ and $y$ of Eq.~(\ref{vecTPs}) can be expressed as
\bea 
a_{\sss V}^\lambda &=& m_{\sss V} g_{\sss V}\left[
\frac{f_{2}}{m_{\Lambda_b}} (X_{\sss V}^{\lambda}+ Y_{\sss
V}^{\lambda}) +f_1\frac{m_{\Lambda_b}-m_{F_1}}{m_{\Lambda_b}^2}
(A_{\sss V}^{\lambda}+ B_{\sss V}^{\lambda}) \right] ~, \nn\\
b_{\sss V}^\lambda &=& m_{\sss V} g_{\sss V} \left[
-\frac{g_{2}}{m_{\Lambda_b}} (X_{\sss V}^{\lambda}- Y_{\sss
V}^{\lambda}) + g_1\frac{m_{\Lambda_b}+m_{F_1}}{m_{\Lambda_b}^2}
(A_{\sss V}^{\Lambda}- B_{\sss V}^{\lambda}) \right] ~, \nn\\
x_{\sss V}^\lambda &=& m_{\sss V} g_{\sss V} [f_{1}-
\frac{m_{\Lambda_b} + m_{\sss F_1}}{m_{\Lambda_b}}f_{2}][X_{\sss
V}^{\lambda}+ Y_{\sss V}^{\lambda}] ~, \nn\\
y_{\sss V}^\lambda &=& -m_{\sss V} g_{\sss V} [g_{1}+
\frac{m_{\Lambda_b}-m_{\sss F_1}}{m_{\Lambda_b}}g_{2}][X_{\sss
V}^{\lambda}- Y_{\sss V}^{\lambda}] ~,
\label{abxyvecgeneral}
\eea
where $\lambda$ denotes the polarization of the final-state $V$, and
we have again dropped subleading terms of $O(m_{\sss V} /
m_{\Lambda_b})$. If any two of the four terms in Eq.~(\ref{Vamps})
above have a relative weak phase, their interference can lead to
triple products.

In the SM, one finds that $A_{\sss V} \simeq B_{\sss V} \approx 0$,
and that $Y_{\sss V} \approx 0$ for a longitudinally-polarized
$V$. For a transversely-polarized $V$, $Y_{\sss V}$ can be nonzero,
but is still quite small. Thus, $\Lambda_b\to F_1 V$ decays are
dominated by a single amplitude (the $X_{\sss V}$ term in
Eq.~(\ref{Vamps}) above), so that triple products in such decays are
expected to be tiny. Specifically, one finds \cite{BDL} that the
triple-product asymmetry in $\Lambda_b \to p K^{*-}$ is $O(1\%)$ for a
transversely-polarized $K^{*-}$, while for a longitudinally-polarized
$K^{*-}$ the asymmetry is $\ll 1\%$. For $\Lambda_b \to \Lambda \phi$
\cite{Kagan} and $\Lambda_b \to n {\bar K}^{*0}$, the asymmetries
essentially vanish since these decays are dominated by a single weak
decay amplitude (the $b\to s$ penguin).

\subsection{$\Lambda_b\to F_1 P$: New Physics}

We begin by considering the new-physics contributions to
triple-product correlations in $\Lambda_b \to p K^-$ decays. Although
this process is governed by the quark transition $b\to s{\bar u}u$,
one still has to perform Fierz transformations on the operators in
Eq.~(\ref{HeffNP}) to put them in a form appropriate for this
decay. Using the relations
\bea
i f_{\sss K} q^{\mu} & = & \bra{K} \bar{s} \gamma^{\mu}(1 - \gamma_5)u
\ket{0} ~, \nn\\
\bra{K} {\bar s} (1 \pm \gamma_5) u \ket{0} & = & \mp {i f_{\sss K}
m_{\sss K}^2 \over m_s + m_u} ~, \nn\\
\bra{p} {\bar u} (1 \pm \gamma_5) b \ket{\Lambda_b} & = & {q^\mu\over
m_b} \bra{p} {\bar u} \gamma_\mu (1 \mp \gamma_5) b \ket{\Lambda_b} ~,
\eea
we find that the new-physics contributions to $X_{\sss K}$ and $Y_{\sss K}$
[Eq.~(\ref{Pfacformulas})] are
\bea
X_{\sss K}^{\sss NP} & = & {G_{\sss F} \over \sqrt{2}} \left[ {1\over 4} a_{u,1}^{\sss
RR} \chi_{\sss K} + {1\over 2} a_{u,1}^{\sss LR} + b_{u,1}^{\sss LL} -
b_{u,1}^{\sss RL} \chi_{\sss K} + 3 c_{u,1}^{\sss RR} \chi_{\sss K}
\right], \nn\\
Y_{\sss K}^{\sss NP} 
& = & {G_{\sss F} \over \sqrt{2}} \left[ - {1\over 4} a_{u,1}^{\sss
LL} \chi_{\sss K} - {1\over 2} a_{u,1}^{\sss RL} - b_{u,1}^{\sss RR} +
b_{u,1}^{\sss LR} \chi_{\sss K} - 3 c_{u,1}^{\sss LL} \chi_{\sss K}
\right],
\label{pK}
\eea
where 
\beq
\chi_{\sss K} \equiv \frac{2 m_{\sss K}^2}{(m_s+m_u)m_b} ~,
\eeq
and we have defined
\beq
a_{q,1}^{\sss AB} \equiv f_{q,1}^{\sss AB} + {1 \over N_c}
f_{q,2}^{\sss AB} ~~,~~ b_{q,1}^{\sss AB} \equiv g_{q,1}^{\sss AB} +
{1 \over N_c} g_{q,2}^{\sss AB} ~~,~~ c_{q,1}^{\sss AB} \equiv
h_{q,1}^{\sss AB} + {1 \over N_c} h_{q,2}^{\sss AB} ~.
\eeq
Note that we can obtain $Y_{\sss K}^{\sss NP}$ from $X_{\sss K}^{\sss
NP}$, up to an overall minus sign, simply by changing the chiralities
$ L \leftrightarrow R$.
 
As discussed in the previous subsection, within the SM the
triple-product correlation for $\Lambda_b \to p K^-$ is expected to be
large, $\sim 18\%$. However, since the new-physics operators of
Eq.~(\ref{HeffNP}) can contribute to both $X_{\sss K}$ and $Y_{\sss
K}$, this prediction can easily be modified.

We now turn to the decay $\Lambda_b \to \Lambda \eta (\eta^{\prime})$,
which receives contributions from all three quark-level processes
$b\to s{\bar q}q$, $q=u,d,s$. The calculation is similar to that
above. For $\Lambda_b \to \Lambda \eta$ we find
\bea
X_\eta^{\sss NP} & = & {G_{\sss F} \over \sqrt{2}} \left[x_u+x_d+x_s+
3 \, c_{s,1}^{\sss RR} \chi_{ \eta_s}\right] , \nn\\
x_{u(d)} & = & r_{u(d)}\left[{1 \over 2} (a_{u(d),2}^{\sss
RL}-a_{u(d),2}^{\sss RR})\chi_{\eta_{u(d)}} +(b_{u(d),2}^{\sss
LL}-b_{u(d),2}^{\sss LR}) \right], \nn\\
x_s & = & r_s\left[{1 \over 2} (a_{s,2}^{\sss RL}-a_{s,2}^{\sss RR}+{1
\over 2}a_{s,1}^{\sss RR} -2b_{s,1}^{\sss RL}) \chi_{\eta_s}
+(b_{s,1}^{\sss LL} + b_{s,2}^{\sss LL} -b_{s,2}^{\sss LR}+{1 \over
2}a_{s,1}^{\sss LR}) \right], \\
Y_\eta^{\sss NP} & = & {G_{\sss F} \over \sqrt{2}} \left[ y_u+y_d+y_s
- 3 \, c_{s,1}^{\sss LL} \chi_{\eta_s}] \right], \nn\\
y_{u(d)} & = & r_{u(d)}\left[{1 \over 2} (-a_{u(d),2}^{\sss
LR}+a_{u(d),2}^{\sss LL}) \chi_{\eta_{u(d)}}+(-b_{u(d),2}^{\sss
RR}+b_{u(d),2}^{\sss RL}) \right], \nn\\
y_s & = & r_s\left[{1 \over 2} (-a_{s,2}^{\sss LR}+a_{s,2}^{\sss
LL}-{1 \over 2}a_{s,1}^{\sss LL} +2b_{s,1}^{\sss LR}) \chi_{\eta_s}
+(-b_{s,1}^{\sss RR} - b_{s,2}^{\sss RR} +b_{s,2}^{\sss RL}-{1 \over
2}a_{s,1}^{\sss RL}) \right], \nn
\label{leta}
\eea
with
\beq
\chi_{\eta_{u,d,s}} = \frac{m_{\eta}^2}{m_{u,d,s} m_b} ~,
\eeq
and
\beq
a_{q,2}^{\sss AB} \equiv f_{q,2}^{\sss AB} + {1 \over N_c}
f_{q,1}^{\sss AB} ~~,~~ b_{q,2}^{\sss AB} \equiv g_{q,2}^{\sss AB} +
{1 \over N_c} g_{q,1}^{\sss AB} ~.
\eeq
In the above, we have defined $r_{u,d,s} \equiv
f_{\eta}^{u,d,s}/f_{\pi}$, where
\bea
if_{\eta}^u p_{\eta}^{\mu} & = &
\bra{\eta}\bar{u}\gamma^{\mu}(1-\gamma_5)u\ket{0}
=\bra{\eta}\bar{d}\gamma^{\mu}(1-\gamma_5)d\ket{0} ~, \nonumber\\
if_{\eta}^sp_{\eta}^{\mu} & = &
\bra{\eta}\bar{s}\gamma^{\mu}(1-\gamma_5)s\ket{0} ~.
\eea
The amplitude for $\Lambda_b \to \Lambda \eta^{\prime}$ has the same
form as Eq.~(\ref{leta}) with the replacement $\eta \to
\eta^{\prime}$. 

Finally, we consider the decay $\Lambda_b \to n {\bar K}^0$, which is
related by isospin to $\Lambda_b \to p K^-$. This is a pure penguin
decay, with $b\to s{\bar d}d$. This decay will be much difficult to
detect experimentally. Nevertheless, we include it here for
completeness. We find
\bea
X_{\bar K}^{\sss NP} & = & {G_{\sss F} \over \sqrt{2}} \left[ {1\over 4}
a_{d,1}^{\sss RR} \chi_{\sss {\bar K}} + {1\over 2} a_{d,1}^{\sss LR}
+ b_{d,1}^{\sss LL} - b_{d,1}^{\sss RL} \chi_{\sss {\bar K}} + 3
c_{d,1}^{\sss RR} \chi_{\sss {\bar K}} \right], \nn\\
Y_{\bar K}^{\sss NP} & = & {G_{\sss F} \over \sqrt{2}} \left[ - {1\over 4}
a_{d,1}^{\sss LL} \chi_{\sss {\bar K}} - {1\over 2} a_{d,1}^{\sss RL}
- b_{d,1}^{\sss RR} + b_{d,1}^{\sss LR} \chi_{\sss {\bar K}} - 3
c_{d,1}^{\sss LL} \chi_{\sss {\bar K}} \right],
\eea
where 
\beq
\chi_{\bar K} \equiv \frac{2 m_{\sss K}^2}{(m_s+m_d)m_b} ~.
\eeq

For each of the decays $\Lambda_b \to \Lambda \eta$, $\Lambda_b \to
\Lambda \eta'$ and $\Lambda_b \to n {\bar K}^0$, the triple product is
tiny in the SM. This is due essentially to the fact that these decays
are dominated by a single weak decay amplitude (the $b\to s$ penguin).
However, this is no longer true in the presence of new physics; on the
contrary, there may be several decay amplitudes. The new-physics
operators may therefore lead to sizeable triple products in these
decays.

\subsection{$\Lambda_b\to F_1 V$: New Physics}

We now examine the new-physics contributions to triple products in
$\Lambda_b\to F_1 V$ decays. Before turning to specific decays, one
can make some very general observations. 

First, the amplitude for the production of a transversely-polarized
vector boson $V$ is suppressed relative to that for a
longitudinally-polarized $V$ by a factor $m_{\sss V}/E_{\sss
V}$. Since $E_{\sss V} \sim m_{\Lambda_b}/2$, this means that this
production amplitude is subleading in the heavy-quark expansion, and
can be neglected. In other words, in our analysis, we will assume the
vector meson in the decay $\Lambda_b\to F_1 V$ to be essentially
longitudinally polarized. As explained earlier, this is justified by
the fact that it is very unlikely that the new physics will affect the
production of a transversely-polarized $V$ without also affecting that
of a longitudinally-polarized $V$.

Second, in the rest frame of the $\Lambda_b$, we can write the
4-momentum of the final state vector meson as $q_\mu = (E_{\sss V}, 0,
0, |\vec{p}_{\sss V}|)$, so that the longitudinal polarization vector
takes the form $\varepsilon_\mu^{\lambda=0} = (1/m_{\sss V})
(|\vec{p}_{\sss V}|, 0, 0, E_{\sss V})$. In the heavy-quark limit,
$E_{\sss V} \gg m_{\sss V}$. Thus, in this limit, the longitudinal
polarization vector can be written approximately as
\beq
\varepsilon_\mu^{\lambda=0} \simeq {1 \over m_{\sss V}} \left( q_\mu +
{m_{\sss V}^2 \over 2 E_{\sss V}} n_\mu \right),
\label{pol}
\eeq
with $n_\mu = (-1,0,0,1)$. In other words, to leading order in the
heavy-quark expansion, $\varepsilon_\mu^{\lambda=0}$ is proportional
to $q_\mu$. This has two important consequences.

Consider first the $A_{\sss V}$ amplitude of Eq.~(\ref{Vamps}), which
is one of the four amplitudes describing $\Lambda_b\to F_1 V$ decays:
\beq
m_{\sss V} g_{\sss V} \varepsilon \cdot (p_{\Lambda_b} + p_{F_1}) \, q_\mu \,
\bra{F_1}\bar{q}_1 \gamma^\mu(1-\gamma_5) b\ket{\Lambda_b} {A_{\sss V}
\over m_{\Lambda_b}^2} ~.
\eeq
Since $p_{\sss F_1}^\mu = (E_{\sss F_1},0,0,-|\vec{p}|)$, one sees
that $\varepsilon^*_{\sss V} \cdot (p_{\Lambda_b} + p_{\sss F_1})$
will be nonzero only for a longitudinally-polarized $V$. Now, writing
the quark content of the $V$ as ${\bar q}_2 q_3$, the operators which
correspond to the $V$ take the form $\bar{q}_2 (1 \pm \gamma_5) q_3$,
$\bar{q}_2 \gamma^{\mu}(1 \pm \gamma_5) q_3$ or $\bar{q}_2
\sigma^{\mu\nu}(1 \pm \gamma_5) q_3$. In calculating the $V$ matrix
elements, these yield
\bea
\bra{V} \bar{q}_2 (1 \pm \gamma_5) q_3 \ket{0} & = & 0 ~, \nn\\
\bra{V} \bar{q}_2 \gamma_{\mu} (1 \pm \gamma_5) q_3 \ket{0} & = & m_{\sss V}
g_{\sss V} \varepsilon_{\mu}^{*} ~, \nn\\
\bra{V} \bar{q}_2 \sigma_{\mu \nu} q_3 \ket{0} & = & -i g_{V}^{\sss T}
\left[ \varepsilon_\mu^* q_\nu - \varepsilon_\nu^* q_\mu \right] ~.
\label{matrixels}
\eea
Thus, we see that it is only the tensor matrix element which could
potentially contribute to $A_{\sss V}$. However, to leading order in
the heavy-quark expansion, $\varepsilon_\mu^{\lambda=0} \sim q_\mu$,
so that the tensor matrix element vanishes. Thus, we have $A_{\sss V}
= O(m_{\sss V}/m_{\Lambda_b})$, even in the presence of new-physics
operators, and we neglect it. This argument applies also to the
$B_{\sss V}$ amplitude of Eq.~(\ref{Vamps}). (By comparison, $X_{\sss
V}$ and $Y_{\sss V}$ are expected to be $O(1)$, i.e.\ leading order in
the heavy-quark expansion.)

Neglecting the $A_{\sss V}$ and $B_{\sss V}$ terms,
Eq.~(\ref{abxyvecgeneral}) reduces to
\bea
a_{\sss V}^\lambda &=& m_{\sss V} g_{\sss V}
\frac{f_{2}}{m_{\Lambda_b}} [X_{\sss V}^{\lambda}+ Y_{\sss
V}^{\lambda}] ~, \nn\\
b_{\sss V}^\lambda &=& -m_{\sss V} g_{\sss V}\frac{g_{2}}{m_{\Lambda_b}}
[X_{\sss V}^{\lambda}- Y_{\sss V}^{\lambda}] ~, \nn\\
x_{\sss V}^\lambda &=& m_{\sss V} g_{\sss V} [f_{1}-
\frac{m_{\sss F_1}+m_{\Lambda_b}}{m_{\Lambda_b}}f_{2}][X_{\sss V}^{\lambda}+
Y_{\sss V}^{\lambda}] ~, \nn\\
y_{\sss V}^\lambda &=& -m_{\sss V} g_{\sss V} [g_{1}+
\frac{m_{\Lambda_b}-m_{\sss F_1}}{m_{\Lambda_b}}g_{2}][X_{\sss V}^{\lambda}-
Y_{\sss V}^{\lambda}] ~.
\label{abxyvec}
\eea
Note that $a_{\sss V}^\lambda$ and $x_{\sss V}^\lambda$ now have the
same weak phase, as do $b_{\sss V}^\lambda$ and $y_{\sss V}^\lambda$.

Now consider again the triple-product terms of Eq.~(\ref{vecTPs}). As
discussed above, to leading order in the heavy-quark expansion, only
longitudinally-polarized vector mesons need be considered, and
$\varepsilon_\mu^{\lambda=0} \sim q_\mu$ in this limit. Thus, we see
that triple products of the form $\epsilon_{\alpha\beta\mu\nu} p_{\sss
F_1}^\alpha p_{\Lambda_b}^\beta s_{\Lambda_b}^\mu \varepsilon_{\sss
V}^\nu$ are of subleading order, and we neglect them. In fact, to
leading order, there is only a single triple product which remains:
\bea
|{\cal M}_{\sss V}|^2_{t.p.} & \simeq & {4 \over m_{\sss V}^2}
\epsilon_{\alpha\beta\mu\nu} p_{\Lambda_b}^\alpha s_{\Lambda_b}^\beta
q^\mu s_{\sss F_1}^\nu \quad \left\{ -2 {\rm Im}(a b^*) \left\vert q
\cdot p_{\Lambda_b} \right\vert^2 + {\rm Im} \left(x y^*\right) q
\cdot p_{\Lambda_b} \right. \nn\\
& & \quad \left. + q \cdot p_{\Lambda_b} \left[ (m_{\Lambda_b} +
m_{\sss F_1}) {\rm Im} \left( a y^* \right) + (m_{\Lambda_b} - m_{\sss
F_1}) {\rm Im} \left( b x^* \right) \right] \right\} ~.
\eea
All other triple products are expected to be smaller, by a factor of
order $m_{\sss V}/m_{\Lambda_b}$.

We now turn to specific decays, and start with $\Lambda_b \to p
K^{*-}$. First, for the tensor operators, one needs to evaluate matrix
elements of the form
\beq
\bra{V} \bar{q}_2 \sigma_{\mu \nu}(1\pm \gamma_5) q_3 \ket{0} 
\bra{F} \bar{q}_3 \sigma_{\mu \nu}(1\pm \gamma_5) b \ket{\Lambda_b} ~.
\eeq
However, as we have argued above, the tensor matrix element vanishes
to leading order in the heavy-quark expansion. Therefore the tensor
operators will not contribute to this decay. The same does not hold
true for the scalar/pseudoscalar and vector/axial vector new-physics
operators, and we find
\bea
X_{\sss K^*}^{\sss NP, \lambda} & = & {G_{\sss F} \over \sqrt{2}} \left[
 - {1\over 2} a_{u,1}^{\sss
LR}  + b_{u,1}^{\sss LL} 
\right], \nn\\
Y_{\sss K^*}^{\sss NP, \lambda} & = & {G_{\sss F} \over \sqrt{2}} \left[
 - {1\over 2} a_{u,1}^{\sss
RL}  + b_{u,1}^{\sss RR}
\right].
\label{pK*NP}
\eea
Note that, as expected, the new-physics operators contribute equally
to longitudinally- and transversely-polarized $V$'s. It is therefore
reasonable to concentrate on the longitudinal $V$'s, which dominate
the decay $\Lambda_b \to p K^{*-}$.

The expressions for the decay $\Lambda_b \to n {\bar{K^{0*}}}$ can be
easily obtained from those above by the replacement $u \to d$.

Finally, for $\Lambda_b \to \Lambda\phi$, we have
\bea
X_{\sss \phi}^{\sss NP, \lambda} & = & {G_{\sss F} \over \sqrt{2}} \left[
 - {1\over 2} a_{s,1}^{\sss
LR}  + b_{s,1}^{\sss LL}+ b_{s,2}^{\sss LL}+ b_{s,2}^{\sss LR} 
\right], \nn\\
X_{\sss \phi}^{\sss SM, \lambda} & = & -{G_{\sss F} \over \sqrt{2}}
 V_{tb}V_{ts}^*\left[ a_3^t+ a_4^t + a_5^t -\frac{1}{2}a_7^t
 -\frac{1}{2}a_9^t -\frac{1}{2}a_{10}^t \right. \nn\\ 
& & \hskip20truemm \left.  -a_3^c- a_4^c - a_5^c +\frac{1}{2}a_7^c
+\frac{1}{2}a_9^c +\frac{1}{2}a_{10}^c \right], \nn\\
Y_{\sss \phi}^{\sss NP, \lambda} & = & {G_{\sss F} \over \sqrt{2}}
\left[ - {1\over 2} a_{s,1}^{\sss RL} + b_{s,1}^{\sss RR}+
b_{s,2}^{\sss RR}+ b_{s,2}^{\sss RL} \right], \nn\\
Y_{\sss \phi}^{\sss SM, \lambda} & \simeq & 0 ~,
\eea
where we have included the standard model contribution without the
tiny dipole contribution. The definitions of the various coefficients
$a_i^q$, as well as their values, can be found in Ref.~\cite{BDL}.

In all of the above decays, $Y_{\sss V}$ is expected to be very small
in the SM, so that the triple products in $\Lambda_b \to F_1 V$ are
predicted to be at most $O(1\%)$. However, this can change
significantly in the presence of new physics -- from the above
expressions one sees that the new-physics operators can easily produce
a nonzero $Y_{\sss V}$. The triple products in $\Lambda_b \to F_1 V$
may well be sizeable in the presence of new physics.

\section{Diagnostic Power}

In the previous section, we saw that the presence of new-physics
operators can significantly modify the SM predictions for
triple-product correlations in $b\to s$ $\Lambda_b$ decays. In
particular, triple products which were expected to be tiny in the SM
may now be sizeable. This is not at all surprising: most of those
triple products are vanishingly small because the decays are dominated
by a single weak $b\to s$ penguin decay amplitude. However, in the
presence of new physics, one can have several decay amplitudes and,
consequently, large triple-product asymmetries.

Although this particular result is entirely expected, the previous
exercise is still useful for several reasons. First, the pattern of
nonzero triple products provides information about the type of
new-physics operators which may be present. And second, one can apply
the above general analysis to specific models of new physics to obtain
model-dependent predictions. These are the issues we explore in this
section.

We begin with the model-independent analysis. The first observation is
simple: if one sees no new effect in a particular decay, this implies
that certain new-physics operators are absent (barring fine-tuned
cancellations among these operators). For example, suppose that the
triple-product asymmetry in $\Lambda_b \to p K^{*-}$ is found to be
tiny, as in the SM. This means that $Y_{\sss K^*}^{\sss NP} = 0$
[Eq.~(\ref{pK*NP})], so that $a_{u,1}^{\sss RL} = b_{u,1}^{\sss RR} =
0$. (Note: since $Y_{\sss K^*}^{\sss SM} \simeq 0$, $X_{\sss
K^*}^{\sss NP}$ could still be nonzero, since the triple product is
proportional to the product of these two quantities.) This in turn
suggests that each of $f_{u,1}^{\sss RL}$, $f_{u,2}^{\sss RL}$,
$g_{u,1}^{\sss RR}$ and $g_{u,2}^{\sss RR}$ vanish, since they make up
$a_{u,1}^{\sss RL}$ and $b_{u,1}^{\sss RR}$. Similarly, should no new
effects be seen in $\Lambda_b \to \Lambda \eta$, each of the 30
operators in $Y_\eta^{\sss NP}$ [Eq.~(\ref{leta})] must vanish.

Of course, one can obtain more information by combining measurements,
since the same operators can contribute to more than one decay. In
fact, one can even partially test the assumption that there are no
fine-tuned cancellations. For example, suppose that the triple-product
asymmetry in $\Lambda_b \to p K^-$ is found to agree with the SM, but
that in $\Lambda_b \to p K^{*-}$ does not. The latter result implies
that $a_{u,1}^{\sss RL}$ and/or $b_{u,1}^{\sss RR}$ are nonzero.
However, these operators also contribute to $Y_{\sss K}^{\sss NP}$
[Eq.~(\ref{pK})]. Thus, in order to obtain $Y_{\sss K}^{\sss NP} = 0$,
there must be cancellations among the various operators. Should such a
result be found, it would be necessary to explain these cancellations,
either via a symmetry, or by construction within a given model.

We now turn to the model-dependent analysis. The very general results
of the previous section can be applied to specific models of new
physics. Of course, in a given model, not all the operators of
Eq.~(\ref{HeffNP}) will appear. In addition, it may be that the
coefficients of those operators which do appear are related in some
way. As examples of this behaviour, we examine those models described
in Sec.~2, but this analysis can be applied to any models of new
physics (e.g.\ supersymmetry, left-right symmetric models, etc.).

Consider first supersymmetric models with R-parity breaking
(Sec.~2.1). If only B-violating couplings are present, then the only
new-physics operators are vector operators contributing to $b\to
s{\bar d}d$ [Eq.~(\ref{Bsdd})]. This leads to a clear pattern of
predictions: no new-physics effects are expected in the decays of a
$\Lambda_b$ to $p K^-$, $p K^{*-}$ and $\Lambda \phi$. Indeed, if
measurements of these triple-product asymmetries disagree with the SM
predictions, this particular model is ruled out. 

On the other hand, the decays $\Lambda_b \to \Lambda \eta$, $n {\bar
K}^0$ and $n {\bar K}^{*0}$ can be affected in this model. How big can
these effects be? In general, they can be enormous. As we have already
noted, the new-physics contributions to these rare decays are still
allowed by data to be comparable to, or even larger than, the SM
contributions. If the two interfering amplitudes are of similar size,
the triple-product asymmetry can be as large as $\sim 50\%$ (to be
contrasted with the SM prediction of $\simeq 0$). This also holds for
the other models discussed below.

Turning to the L-violating couplings, one sees that more operators may
be present [Eqs.~(\ref{BsuuL}), (\ref{BsddL}), (\ref{BsssL})]. In this
case, all decays may be affected, except $\Lambda_b$ to $p K^-$. This
is a a quite distinctive signature for this model.

Finally, we consider leptophobic $Z'$-mediated FCNC's (Sec.~2.2).
There are only six nonzero new-physics coefficients, given in
Eq.~(\ref{Z'FCNCcoeffs}), and these all depend on the parameters
$\left| U^{\sss Z'}_{sb} \right|$ and $M_{\sss Z'}$
[Eq.~(\ref{Z'constraint2})]. In this case, all $\Lambda_b$ decays will
be affected. However, note that, within this model, there are more
observables (6) than there are theoretical parameters (2). This means
that if deviations from the SM predictions are measured, we will be
able to get a handle on $\left| U^{\sss Z'}_{sb} \right|$ and $M_{\sss
Z'}$. Conversely, if no new-physics effects are observed, we will be
able to place strong constraints on these quantities.

\section{Conclusions}

In the standard model (SM), (almost) all T-violating triple-product
correlations in charmless $\Lambda_b$ decays are expected to be tiny.
(The one exception is the decay $\Lambda_b \to p K^-$, for which the
asymmetry is 18\%.) This is therefore a good place to look for physics
beyond the standard model.

In this paper, using an effective-lagrangian approach, we have
computed the effects of new physics on such triple products. This
approach has the advantage of indicating which specific new-physics
operators affect each of the $\Lambda_b$ triple-product correlations.
Thus, the measurement of a number of different triple products permits
us to determine which new-physics operators are or are not present.
Furthermore, the approach is completely general -- the effects of any
specific model can be obtained by simply calculating which operators
appear in that model.

The new-physics effects on triple products are calculated using
factorization. In addition, we work only to leading order in the
heavy-quark expansion, neglecting terms of order $m/m_{\Lambda_b}$,
where $m$ is the mass of the light final-state meson. The
justification for this is that it is only in fine-tuned scenarios that
the new physics contributes at subleading order, but not at leading
order. (Also, the subleading terms are quite a bit smaller, e.g.\
$m_{K^*}/m_{\Lambda_b} \sim 15\%$.)

We have found that all $\Lambda_b$ triple products can be
significantly modified by new physics. Of course, this to be expected.
Most of the triple products are vanishingly small in the SM because
the decays are dominated by a single weak $b\to s$ penguin decay
amplitude. Thus, in the presence of new physics, there may be several
decay amplitudes which can interfere with the SM amplitude. However,
in order to obtain a sizeable asymmetry, the interfering amplitudes
must be of similar size. We note that the constraints on the
new-physics operators are sufficiently weak that they can be
comparable to, or even larger than, the SM contributions. Thus, triple
products which vanish in the SM can be as large as $\sim 50\%$ with
new physics.

We have demonstrated how the measurement of triple-product asymmetries
provides diagnostic information about the new-physics operators
present. For example, all operators which affect $\Lambda_b \to p
K^{*-}$ also affect $\Lambda_b \to p K^-$, but not vice-versa. Thus,
if the triple product in $\Lambda_b \to p K^-$ is found to agree with
the SM, we would also expect no new effects in $\Lambda_b \to p
K^{*-}$. If this were found not to hold, then we would conclude that
there must be cancellations among the operators in $\Lambda_b \to p
K^-$, and this would have to be explained in some way (e.g.\ symmetry,
specific model, etc.). 

Finally, we have also applied this general approach to two specific
models: supersymmetry with R-parity breaking, and leptophobic
$Z'$-mediated flavour-changing neutral currents. In both cases, we
have worked out the new-physics operators which appear in those
models, and used the previous formalism to calculate which $\Lambda_b$
triple products can be affected. For example, in the case of R-parity
breaking models, there is a clear pattern of effects. One such model
predicts significant new effects in the decays $\Lambda_b \to \Lambda
\eta$, $n {\bar K}^0$ and $n {\bar K}^{*0}$, but not in $\Lambda_b$ to
$p K^-$, $p K^{*-}$ and $\Lambda \phi$. Any deviation from this
pattern would rule out this model. Other models of new physics can be
treated similarly.

\bigskip
\noindent
{\bf Acknowledgements}:
This work was financially supported by NSERC of Canada.



\begin{thebibliography}{99}

\bibitem{CPreview} For a review, see, for example, {\it The BaBar
    Physics Book}, eds.\ P.F. Harrison and H.R. Quinn, SLAC Report
  504, October 1998.

\bibitem{betameas} B. Aubert {\it et al.}  [BABAR Collaboration],
arXiv:hep-ex/0203007; K. Trabelsi [BELLE Collaboration], talk given at
the {\it $XXXVII^{th}$ Rencontres de Moriond Electroweak Interactions
and Unified Theories}, 2002.

\bibitem{Valencia} Triple products in $B \to V_1 V_2$ decays have been
studied in G. Valencia, \prd{39}{1989}{3339}.

\bibitem{BenLon} For a discussion of triple products at the quark
level, see W. Bensalem and D. London, \newprd{64}{2001}{116003}.

\bibitem{BDL} W. Bensalem, A. Datta and D. London,
\plb{538}{2002}{309}.

\bibitem{Bnewphysics} For a discussion of the effects of these
new-physics models on $B$ decays, see M. Gronau and D. London,
\prd{55}{1997}{2845}.
 
\bibitem{Proton} I. Hinchliffe and T. Kaeding, \prd{47}{1993}{279};
C.E. Carlson, P. Roy and M. Sher, \plb{357}{1995}{99}; A.Y. Smirnov
and F. Vissani, \plb{380}{1996}{317}.

\bibitem{Rpreview} For recent reviews on $R$-parity violation see
G. Bhattacharyya, arXiv:hep-ph/9709395 and references therein;
H. Dreiner, {\it An Introduction to Explicit R-Parity Violation} in
{\it Perspectives on Supersymmetry}, p.462-479, Ed. G.L. Kane (World
Scientific) and references therein, arXiv:hep-ph/9707435; R. Barbier
{\it et al}, {\it Report of the Group on the $R$-parity Violation},
hep-ph/9810232 (unpublished); B.C. Allanach, A. Dedes and
H.K. Dreiner, \newprd{60}{1999}{075014}, and references therein.
 
\bibitem{He} D.K. Ghosh, X.G. He, B.H. McKellar and J.Q. Shi,
arXiv:hep-ph/0111106.

\bibitem{DatXin} See for example A. Datta, J.M. Yang, B.L. Young and
X. Zhang, \prd{56}{1997}{3107}.

\bibitem{NirSilv} Y. Nir and D. Silverman, \prd{42}{1990}{1477}.

\bibitem{BELLE} K. Abe et al., (BELLE Collaboration), hep-ex/0107072.

\bibitem{Z'FCNC} E. Nardi, \prd{48}{1993}{1240}; J. Bernab\'eu,
E. Nardi and D. Tommasini, \npb{409}{1993}{69}.
  
\bibitem{LerLon} K. Leroux and D. London, \plb{526}{2002}{97}.

\bibitem{bsnunu} Y. Grossman, Z. Ligeti and E. Nardi,
\npb{465}{1996}{753}, (Erratum) \npb{480}{1996}{753}.

\bibitem{nonfacdecay} K. Hagiwara et al.\ (Particle Data Group),
  \newprd{66}{2002}{010001}, pg.\ 66.

\bibitem{Cheng} H.Y. Cheng, \plb{289}{1992}{455}.

\bibitem{SPT} S. Pakvasa, S.P. Rosen and S.F. Tuan,
\prd{42}{1990}{3746}.

\bibitem{Kagan} The usefulness of the decay $\Lambda_b \to
\Lambda\phi$ for probing non-SM-chirality penguins has been discussed
in G. Hiller and A. Kagan, Phys.\ Rev.\ D {\bf 65}, 074038 (2002), and
Z.g. Zhao {\it et al.}, in {\it Proc. of the APS/DPF/DPB Summer Study
on the Future of Particle Physics (Snowmass 2001) } ed. R.~Davidson
and C.~Quigg.

\end{thebibliography}
\end{document}